\documentclass[10pt]{iopart}
\pdfoutput=1
\usepackage{graphicx}
\usepackage{amssymb}
\usepackage{iopams}

\eqnobysec
\usepackage{iopams} 
\usepackage[colorlinks,linkcolor=blue,urlcolor=black,citecolor=blue]{hyperref}

\newcommand{\beq}{\begin{equation}}
\newcommand{\eeq}{\end{equation}}
\newcommand{\be}{\begin{equation*}}
\newcommand{\ee}{\end{equation*}}
\newcommand{\beqa}{\begin{eqnarray}}
\newcommand{\eeqa}{\end{eqnarray}}
\newcommand{\bea}{\begin{eqnarray*}}
\newcommand{\eea}{\end{eqnarray*}}

\def\stackunder#1#2{\mathrel{\mathop{#1}\limits_{#2}}}
\def\binom#1#2{{#1\choose #2}}

\newcommand{\mean}[1]{\langle#1\rangle}
\newcommand{\bigac}[1]{\left(#1\right)}
\newcommand{\lap}[1]{\mathrel{\mathop{\cal L}\limits_{#1}^{}}}

\newcommand{\prob}{\mathbb{P}}
\newcommand{\dd}{{\rm d}}
\newcommand{\cc}{{\cal C}}
\newcommand{\eq}{{\rm eq}}

\renewcommand{\th}{{\theta}}

\newcommand{\xm}{\langle X\rangle}
\newcommand{\num}{_{|{\rm num}}}
\newcommand{\bb}{{\rm bridge}}
\newcommand{\fo}{\ell}

\begin{document}

\title[The Buffon needle problem for L\'evy distributed spacings]{The Buffon needle problem for L\'evy distributed spacings and renewal theory}
\author{Claude Godr\`eche}
\address{
Universit\'e Paris-Saclay, CNRS, CEA,
Institut de Physique Th\'eorique,
91191 Gif-sur-Yvette, France}
\smallskip

\begin{abstract}
What is the probability that a needle dropped at random on a set of points scattered on a line segment does not fall on any of them?
We compute the exact scaling expression of this hole probability when the spacings between the points are independent identically distributed random variables with 
a power-law distribution of index less than unity, implying that the average spacing diverges.
The theoretical framework for such a setting is renewal theory, to which the present study brings a new contribution.
The question posed here is also related to the study of some correlation functions of simple models of statistical physics.

\end{abstract}

\eads{\mailto{claude.godreche@ipht.fr}}

\section{Introduction}

In the classical geometric probability problem devised by Buffon, a needle of length $r$ is dropped at random on a plane on which a set of parallel lines separated by a distance $d$ ($d>r$) have been drawn. 
As is well known, the probability that the needle does not intersect a line
 is equal to $1-(2r)/(\pi d)$, where
the geometrical factor $2/\pi$ comes from the possible orientations of the needle \cite{kendall,feller,calka}.
This factor is no longer present in the one-dimensional Buffon problem where the lines are replaced by equidistant points, in which case the probability that the needle does not hit a point simplifies to
$1-r/d$.

\begin{figure}[!ht]
\begin{center}
\includegraphics[angle=0,width=.9\linewidth]{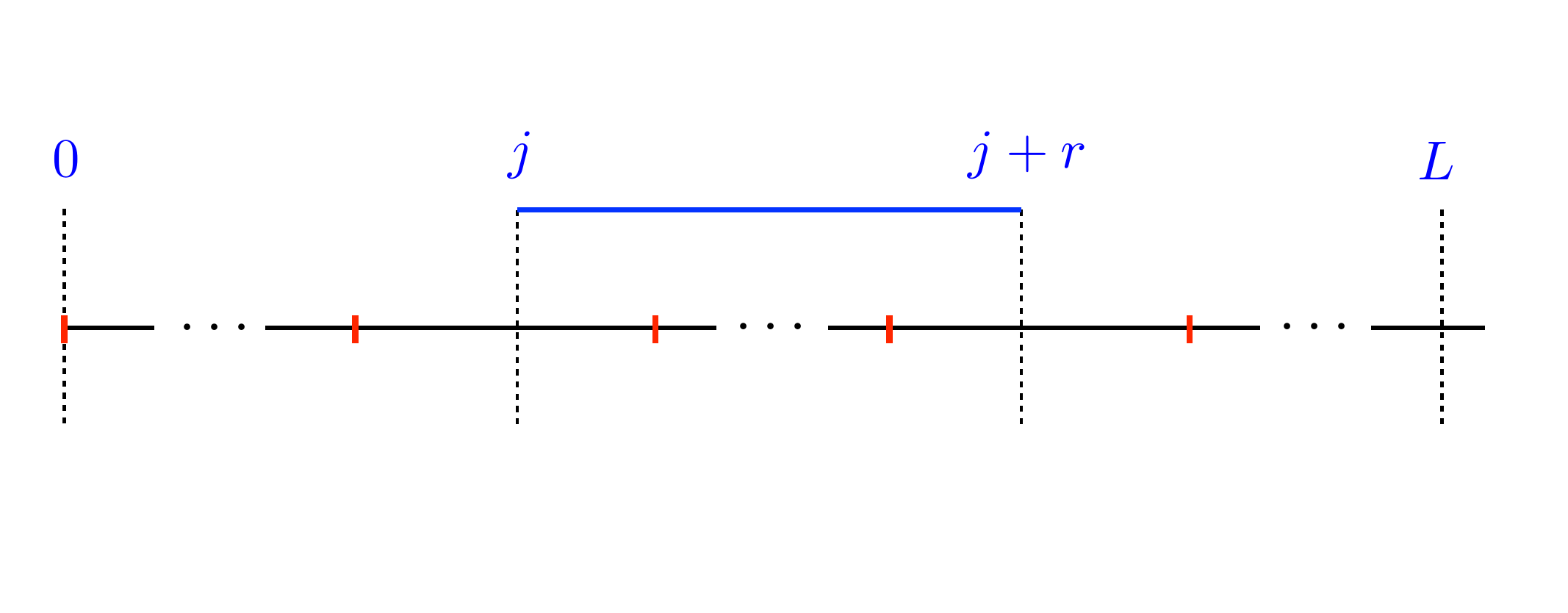}
\caption
{
Setting for a free renewal process.
Point events (or renewals) are figured by red ticks on the (spatial or temporal) coordinate axis.
Spacings between them are iid random variables.
A needle of size $r$ 
is dropped at random on the semi-infinite line.
The first question is to determine the chance for the needle, with left endpoint located at $j$, not to cover a point, i.e., for the interval $(j,j+r)$ to be empty.
Transcribed in the temporal domain, the question is to determine the chance for a customer arriving at time $j$ to wait more than $r$ units of time before getting a taxi.
Solving the Buffon problem then consists in determining the same hole probability when the interval of length $r$ is placed uniformly at random on the line.
In order to give a precise meaning to this question, the system is put in a box of size $L$.
}
\label{fig:definition-libre}
\end{center}
\end{figure}

\begin{figure}[!ht]
\begin{center}
\includegraphics[angle=0,width=.9\linewidth]{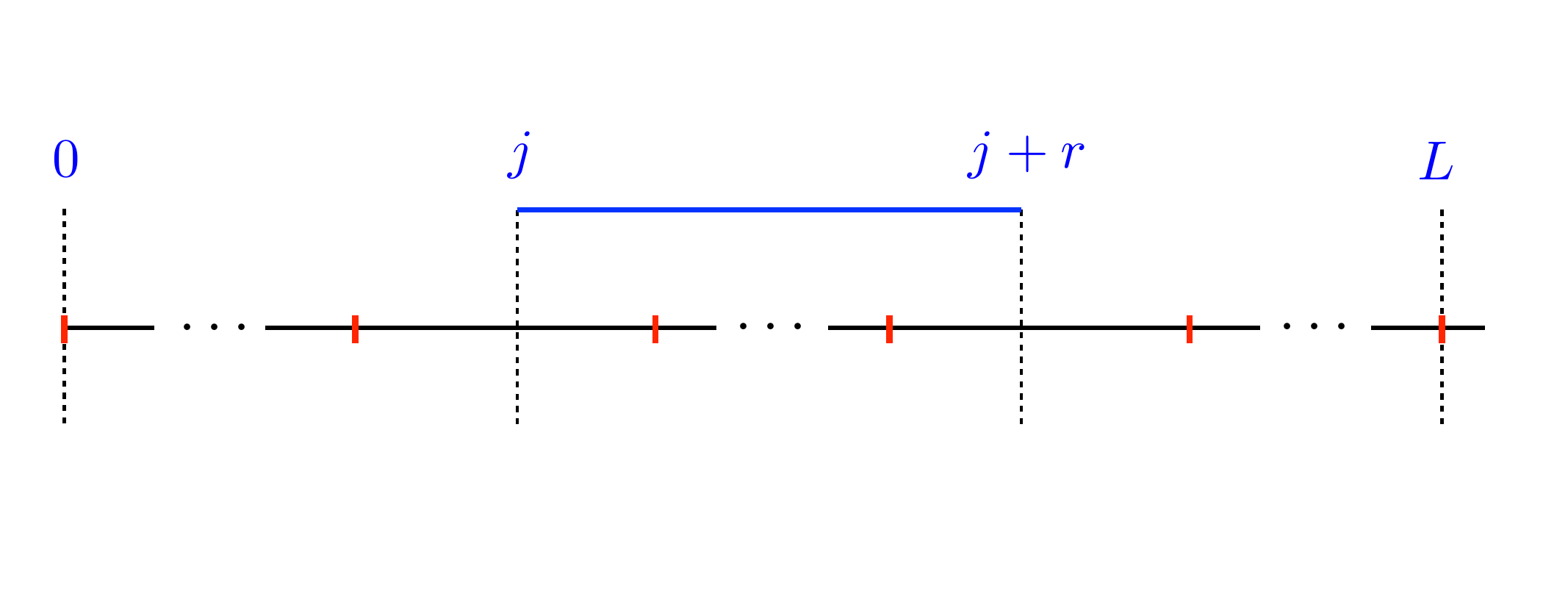}
\caption
{
Setting for a tied-down renewal process.
The same questions can be posed if the configurations in figure \ref{fig:definition-libre} are conditioned
by the presence of a renewal at $L$, or in other words, if the segment $(0,L)$ is partitioned by a random number of intervals $X_1,X_2,\ldots$
}
\label{fig:definition-TD}
\end{center}
\end{figure}

The goal of the present work is to generalise the one-dimensional Buffon problem with equidistant points to the situation where the points are randomly spaced---the spacings between the points are independent, identically distributed (iid) random variables---with particular focus on the case where their common distribution is heavy-tailed, with tail index $\th<1$.
The question posed is: what is the chance for an interval of length $r$ (representing the needle) dropped at random on this set of points not to cover any point
(see figures \ref{fig:definition-libre} and \ref{fig:definition-TD}), in short, what is the probability of a point-free interval of length $r$?
As will be explained below, this question naturally arises in some particular models of statistical physics.

This very same question can be translated in the temporal domain.
A customer is waiting for the arrival of a taxi at the airport.
What is his chance of waiting for more than $r$ units of time before he gets one?

A prerequisite for answering this question
consists in determining the chance for the needle, with left endpoint located at $j$, not to cover a point, i.e., for the interval $(j,j+r)$ to be empty.
Transcribed in the temporal domain, the question is to determine the chance for a customer arriving at time $j$ to wait more than $r$ units of time before getting a taxi.
Solving the Buffon problem then consists in determining the same hole probability when the interval of length $r$ is placed uniformly at random on the line (see figures \ref{fig:definition-libre} and \ref{fig:definition-TD}).
 
Whenever the distribution of spacings between the points equilibrates at large distance (or large time)---which requires its first moment to be finite---the answers to these questions are simple, as will be recalled below.
However, when this distribution does not possess a finite first moment, i.e., when the average spacing between the points is infinite,
the problem is more subtle.
Studies on this topic can be found in \cite{glrenew} for the \textit{free} process, i.e., infinite to the right, as in figure \ref{fig:definition-libre}, and in \cite{wendel2,arxiv,comment}%
\footnote{The context around \cite{arxiv,comment}, as well as their connections to \cite{bmm1,bmm}, where similar questions are tackled, are discussed in \S \ref{sec:discussion}.}
for the \textit{tied-down} process, conditioned to a fixed (spatial or temporal) length $L$ as in figure \ref{fig:definition-TD}; that is, when
the segment of length $L$ is partitioned by a random number of iid intervals (see figure \ref{fig:definition-TD}).

The present work improves and completes these previous studies.
For free renewal processes, the central result is the universal scaling form of the Buffon probability (\ref{eq:buffonfree}).
For tied-down renewal processes, the central results are
the universal scaling function (\ref{eq:scalingfctn}) from which the Buffon probability (\ref{eq:scaling1}) ensues.
The study made in \cite{wendel2} only predicted the form of the scaling function in the two regimes of interest analysed in \S \ref{sec:scaling} and given by (\ref{eq:Cxyshort}) and (\ref{eq:Cxylarge}).
Together with \cite{comment} the present study supersedes the analysis made in \cite{arxiv}.

The paper is structured as follows.
Section \ref{sec:definition} gives the definitions of the renewal process under study, either with free or tied-down boundary conditions.
Section \ref{sec:p0free} gives a survey of known results on the probability of an empty interval 
for a free renewal process, from which the Buffon probability (\ref{eq:buffonfree}) ensues.
Section \ref{sec:p0TD}, which is the most important section of the present work, 
starts by deriving the general expression of the probability of an empty interval 
for a tied-down renewal process, then illustrates the formalism by examples of systems converging to equilibrium.
The main results are contained in \S \ref{sec:scaling}, which
gives the asymptotic analysis of the probability of an empty interval for distributions with power-law fall-off and tail index $\th<1$, for a tied-down renewal process, while 
 \S \ref{sec:variant} is devoted to a variant of the formalism.
 Section \ref{sec:complement} complements the body of the text.
 Section \ref{sec:discussion} provides a discussion.
Other complements can be found in the two appendices.

%
\section{Definition of the process}
\label{sec:definition}

We consider the following point process 
whose definition can be indifferently given in the temporal or spatial domains.
Events occur at the random epochs of time (or space coordinates) $S_{1},S_{2},\ldots$, from some origin $S_0=0$. 
When the intervals of time (space) between events, $X_{1}=S_{1},X_{2}=S_{2}-S_{1},\ldots $, are independent and identically distributed random variables, the process thus
formed is a \textit{renewal process} \cite{feller,grimmett,cox,coxmiller}.
Hereafter we shall use synonymously
the denominations \textit{events} or \textit{renewals}.
We take the origin of time (space) on a renewal.
The sum $S_n$ therefore reads
\beq\label{eq:Sn}
S_n=X_1+\cdots+X_n.
\eeq

The common distribution of the iid random variables $X_{1},X_{2},\ldots $ is denoted by $f(k)=\prob(X=k)$ when $X$ is discrete, and we keep the notation 
$f(k)$ to designate the density when $X$ is a continuous random variable.
In the sequel we shall use both the discrete and continuum formalisms.
Thus the variables $j,k, r, L$ used below will stand either for integers or for real numbers.
The transcription of one formalism to another is easy.

\begin{figure}[!ht]
\begin{center}
\includegraphics[angle=0,width=.9\linewidth]{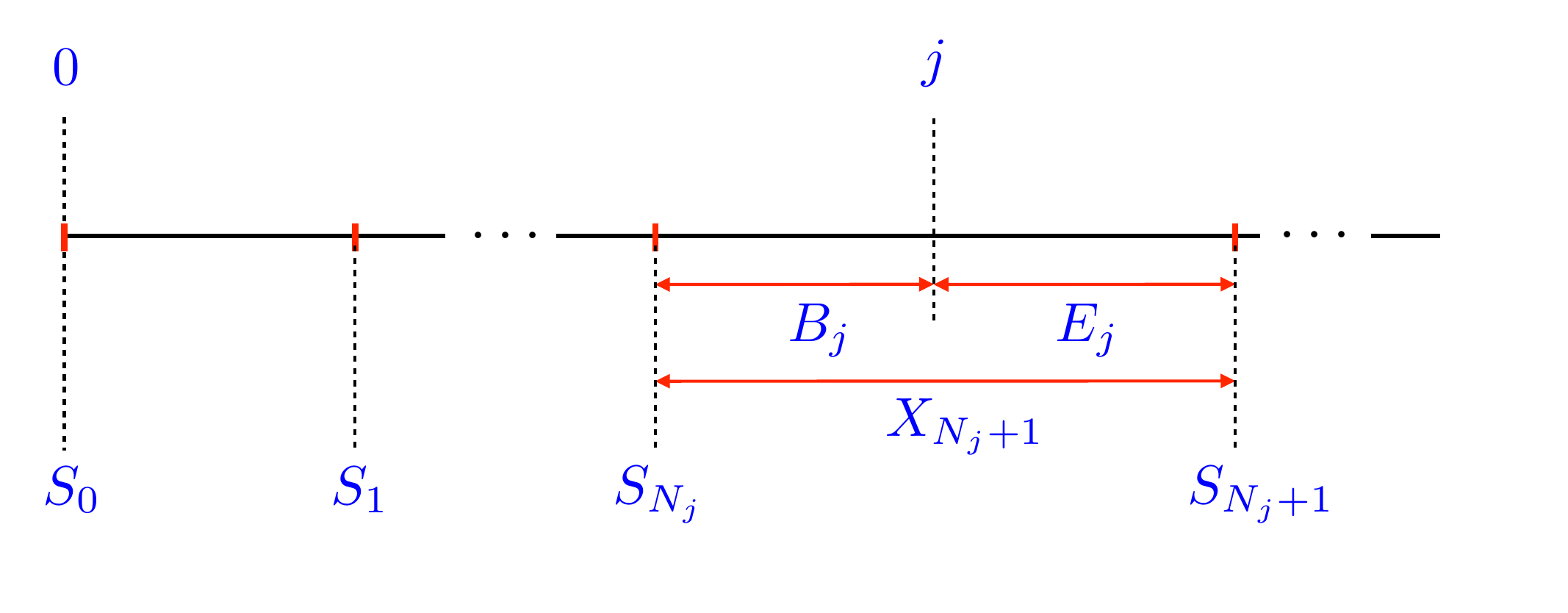}
\caption
{In renewal theory, considering $j$ as a temporal coordinate, $S_{N_j}$ is the epoch of the last event before $j$, 
$S_{N_j+1}$ is the epoch of the first event after $j$, and
$X_{N_j+1}$ is the time interval straddling $j$.
The backward and forward recurrence-times, $B_j=j-S_{N_{j}}$ and $E_j=S_{N_{j}+1}-j$, are respectively the lapses of time between $j$ and the last event, and between 
$j$ and the next event.
}
\label{fig:figEAS}
\end{center}
\end{figure}

The renewal process can be \textit{free}, i.e., unbounded to the right as in figure \ref{fig:definition-libre}, or \textit{tied-down} \cite{wendel2,wendel,wendel1,jsp}, i.e., conditioned by the presence of a renewal at $L$, as in figure \ref{fig:definition-TD}.
\vskip4pt
\noindent
In the first case, denoting by $N_j$ the random number of renewals between 0 and $j$, 
the epoch of the last renewal before $j$ is
\be
S_{N_j}=X_1+\cdots+X_{N_j}.
\ee
The backward recurrence-time $B_j$ and forward recurrence-time $E_j$ (or excess time) are defined respectively as
\beq\label{eq:defEB}
B_j=j-S_{N_j},\qquad E_j=S_{N_j+1}-j,
\eeq
and the interval straddling $j$ is made of their sum,
\beq\label{eq:straddef}
X_{N_j+1}=B_j+E_j=S_{N_j+1}-S_{N_j}.
\eeq
These definitions are illustrated in figure \ref{fig:figEAS}.

\vskip4pt
\noindent
In the second case, i.e., if the intervals $X_i$ are conditioned to sum up to the fixed length $L$, 
we have 
\beq\label{eq:condition}
S_{N_L}=X_1+\cdots+X_{N_L}=L,
\eeq 
i.e., $B_L=0$.
 The simplest realisation of such a process (in the temporal domain) is the simple random walk, with steps $\pm1$, starting and ending at the origin.
The $X_i$ are the time intervals between two returns of the walk at the origin.
This walk is dubbed the \textit{tied-down random walk} because it starts and ends at the origin \cite{wendel,grimmett}, or else the \textit{random walk bridge} (see \S \ref{sec:p0TD}).
Its continuous limit is the \textit{Brownian bridge}, where now the intervals $X_i$ are continuous random variables.

Such systems, made of a random number of iid intervals conditioned by the value of their sum, dubbed \textit{tied-down renewal processes}, are naturally encountered in statistical physics.
The model introduced in \cite{burda2,bar2} is an example, corresponding to a particular choice of distribution $f(k)$ with power-law tail (\ref{eq:powerlaw}), (see \cite{burda2,bar2,bmm,comment,jsp} for details).
In the language of balls in boxes used in \cite{burda2}, the system is made of a random number of boxes, $N_L$, with a fixed total number $L$ of balls in them.
The occupations of the boxes are the lengths of the intervals $X_1,X_2,\dots,X_{N_L}$.
These models belong to the broader class of \textit{linear models} defined by Fisher in \cite{fisher}, such as the Poland-Scherraga model \cite{poland,poland2} or models of wetting \cite{gia1,gia2}.
As explained at the end of this paper, the probability of an empty interval is a key quantity to discuss correlations in such models.

We complete the definition of the process under study by some more details on the distribution
of intervals $f(k)$.
In what follows, $f(k)$ will be either a narrow distribution with finite
moments, or a broad distribution characterised by a
power-law tail with index $\theta $ and parameter $c$, or \textit{L\'evy distribution} for short \cite{bardou},
\beq\label{eq:powerlaw}
f(k)\stackunder{\approx}{k\to\infty} \frac{c}{k^{1+\th}}.
\eeq
If $\theta <1$ all moments of 
$f(k)$ are divergent, while if $1<\theta <2$, the first moment 
$\xm$ is finite but higher moments are
divergent, and so on.
In Laplace space, where $s$ is conjugate to $ k $, for a narrow
distribution we have (see \ref{app:word} for notations)
\beq\label{ro_narrow}
\lap{k} f (k)=\hat{f}(s)=\mean{\e^{-sX}}
\stackunder{=}{s\rightarrow 0}1-\xm s+\frac{1}{2}
\mean{X^2} s^{2}+\cdots 
\eeq
For a broad distribution, we have 
\beq\label{eq:expansionLaplace}
\hat f(s)\stackunder{\approx}{s\to0}\cases{1-|a| s^\th&$\th<1$
\\
1-s\xm +\cdots+a s^\th&$\th>1$,}
\eeq
with
\beq\label{eq:adef}
a=c\,\Gamma(-\th).
\eeq
The parameter $a$ is negative if $0<\theta<1$, positive if $1<\theta<2$, and so on.

\section{Probability of an empty interval for a free renewal process}
\label{sec:p0free}
We start with a brief survey of some known results on the probability of an empty interval 
for a free renewal process (see figure \ref{fig:definition-libre}).
The theory of renewal processes is a classics in probability theory \cite{feller,grimmett,cox,coxmiller}.
The survey presented below relies on the methods of \cite{glrenew}, which makes systematic usage of Laplace transforms.
We give the main structure of the reasoning made in \cite{glrenew}, without repeating all the details of the calculations. 
We then infer the integrated Buffon probability, both at equilibrium and for nonequilibrium distributions with tail index $\th<1$ (see (\ref{eq:buffonfree})).
This material will allow to highlight the resemblances and differences between the free and tied-down processes.

\subsection{Number of renewals between two arbitrary times}

Consider the number of events $N(j,r)=N_{j+r}-N_{j}$
occurring between $j$ and $j+r$, and its probability distribution, denoted as 
\beq\label{eq:pnjr}
p_{n}(j,r)=\prob\left( N(j,r)=n\right) . 
\eeq
The time of occurrence of the $n$-th event, counted from time $j$, is denoted by $T_{n}$, with, by convention, $T_{0}=0$.
By definition of the forward recurrence-time $E_j$ (see (\ref{eq:defEB})), the first event after
time $j$ occurs at time $T_{1}=E_j$, when counted from time $j$.
Hence the time of occurrence of the $n-$th event between $j$ and $j+r$ reads 
\[
T_n=E_j+ X_{2}+\cdots + X_n. 
\]
Therefore, 
\beq\label{pnjr}
p_{n}(j,r)=\prob\left( T_{n}<r
<T_{n+1}\right) .
\eeq
In particular, for $n=0$, we have 
\beq
p_{0}(j,r)=\prob(E_j>r)=\int_{r}^{\infty}\dd \fo\,f_{E}(j,\fo), 
\label{p0jr}
\eeq
where $f_{E}(j,\fo)$ is the density of $E_j$ (see \ref{app:word} for notations).
In other words, the probability of an empty interval $p_{0}(j,r)$ is the survival probability up to time $j+r$, counted from time $j$.

In Laplace space, where $u$ is conjugate to $r$, (\ref{pnjr}) and (\ref{p0jr}) lead to \cite{glrenew}
\begin{eqnarray}
\lap{r} p_{n}(j,r) 
&=&\hat{p}_{n}(j,u)=\hat{f}_{E}(j,u)\,\hat{f}(u)^{n-1}\frac{1-\hat{f}(u)}{u}
\qquad (n\geq 1), 
\nonumber \\
\lap{r}p_{0}(j,r) 
&=&\hat{p}_{0}(j,u)=\frac{1-\hat{f}_{E}(j,u)}{u}, 
\label{pntu}
\end{eqnarray}
where $\hat{f}_{E}(j,u)$ is the Laplace transform of $f_{E}(j,\fo)$ with respect to $\fo$ (see \ref{app:word} for notations).
The Laplace transform of the latter with respect to $j$ reads \cite{glrenew}
\beq\label{fE}
\hat{f}_{E}(s,u) =\frac{\hat{f}(u)-\hat{f}(s)}{s-u}\frac{1}{1-\hat{f}(s)}.
\eeq

The implications of the results above are now discussed according to the nature of the distribution of intervals $f(k)$.

\subsection{Probability of an empty interval at equilibrium}
\label{sec:equilibrium}

If the first moment $\xm$ of the distribution $f(k)$ is finite, i.e., for narrow distributions of intervals or distributions with power-law tail (\ref{eq:powerlaw}) of index $\th>1$,
the process reaches equilibrium at large distances (or long times) \cite{glrenew,cox,coxmiller}, in particular,
\beq\label{eq:limeqp0}
\lim_{j\rightarrow \infty}p_0(j,r)=(p_0)_\eq (r).
\eeq
The answer to the Buffon problem ensues.
Indeed, consider a large (time or space) interval $L$ (see figure \ref{fig:definition-libre}), then the probability of an empty interval placed at random on the line is defined as the limit, when $L\to\infty$ of
\beq\label{eq:P0rLdef}
P_0(r,L)=\frac{1}{L-r}\int_0^{L-r}\dd j\,p_0(j,r).
\eeq
Using (\ref{eq:limeqp0}), we obtain
\be
P_0(r)=\lim_{L\to\infty}P_0(r,L)=(p_0)_\eq (r).
\ee
At equilibrium, the two quantities $(p_0)_\eq(r)$ and $P_0(r)$ are the same.
This holds whether the process is free or conditioned by (\ref{eq:condition}) since both $j$ and $L$ are sent to infinity.

Let us start with the simple example of an exponential
distribution of time (space) intervals with $f(k)=\lambda\,\e^{-\lambda k}$.
Then (\ref{pntu}) and (\ref{fE}) imply that
\be
f_{E}(j,\fo)={\e}^{-\lambda \fo},
\ee
\beq\label{eq:star2}
p_{n}(j,r)=
{\e}^{-\lambda r}\frac{(\lambda r)^{n}}{n!}
\qquad (n\geq 0), 
\eeq
which are independent of $j$, showing that the Poisson point process is at
equilibrium at all times.

The explicit expression of $(p_0)_\eq (r)$ is given in (\ref{eq:equilib}).
In order to derive this expression we proceed as follows.
For the forward recurrence-time, taking the limit of (\ref{fE}) for $s\to0$, we get
\be
(\hat{f}_{E})_\eq(u)=\lim_{s\rightarrow 0}s\hat{f}_{E}(s,u)
=\frac{1-\hat{f}(u)}{\xm u},
\ee
yielding the classical result, 
\beq\label{eq:fEeq}
(f_{E})_\eq(\fo)
=\frac{1}{\xm}\int_{\fo}^{\infty}\,\dd k \,f(k).
\eeq
Therefore, from (\ref{pntu}),
\beqa\label{eq:circ}
(\hat p_{n})_\eq(u)=(\hat{f}_{E})_\eq(u)\,\hat{f}(u)^{n-1}\frac{1-\hat{f}(u)}{u}
\qquad (n\ge1),
\\
\label{eq:star}
(\hat p_0)_\eq(u)=\frac{1-(\hat{f}_{E})_\eq(u)}{u}.
\eeqa
Thus, at equilibrium, the distribution of the random variable $N(j,r)$ no longer depends on $j$. In particular, its average, $\mean{N(j,r)}$ is equal to $r/\xm$.
By inversion of (\ref{eq:star}), we get
\be
(p_0)_\eq(r)=\int_{r}^\infty\dd \fo\,(f_{E})_\eq(\fo)
=\int_{r}^\infty\dd \fo\,
\frac{1}{\xm}\int_{\fo}^{\infty}\,\dd k \,f(k).
\ee
Changing the order of the integrations, one finally finds
\beq\label{eq:equilib}
(p_0)_\eq(r)
=\frac{1}{\xm}\int_{r}^\infty\dd k \,f(k)( k -r).
\eeq

This result can be recovered by the following alternative reasoning.
We first note that, at equilibrium, the density of $X_{N_j+1}$, the interval straddling $j$ (see figure \ref{fig:figEAS}), reads 
\beq\label{eq:straddle}
(f_{X_{N_j+1}})_{\eq}(k)=\frac{k\,f(k)}{\xm}.
\eeq
The result (\ref{eq:straddle}) has a simple derivation (see, e.g., \S \ref{sec:stradl}).
This distribution is dubbed the \textit{length-biased sampling} of $f(k)$ \cite{cox}.
Its interpretation is that if one samples intervals from the distribution $f(k)$, the probability of selecting any one of them is proportional to its size.
The fact that the distribution of the interval straddling $j$ is not the same as the distribution $f(k)$ of a generic interval $X_1,X_2,\ldots$, 
entails the \textit{inspection paradox} \cite{feller} (see \S \ref{sec:stradl}).
Now, taking the average, with this distribution, of the ratio of the available space for the needle to the total length interval,
we infer that
\beq\label{eq:p0eqq}
(p_0)_\eq(r)=\int_{r}^\infty\dd k\,(f_{X_{N_j+1}})_{\eq}(k)\, \frac{k-r}{k}.
\eeq
Using now (\ref{eq:straddle}) in (\ref{eq:p0eqq}) reproduces (\ref{eq:equilib}).

The transcription of these expressions for discrete random variables is straightforward, in particular,
\beq\label{eq:discretEquil}
(p_{0})_{\eq}(r)=\frac{1}{\xm}\sum_{k\ge r}f(k)(k-r).
\eeq

Let us illustrate the preceding analysis on simple examples.

\vskip4pt\noindent\textbf{(a)}
Coming back to the exponential distribution $f(k)=\lambda\,\e^{-\lambda k}$, and using (\ref{eq:equilib}) we instantly get
\be
(p_0)_\eq(r)=\e^{-\lambda r},
\ee
in accord with (\ref{eq:star2}).
We note that, for microscopic values of $r$, we have
\beq\label{eq:micror}
(p_0)_\eq(r)\stackunder{\approx}{r\to0}1-\frac{r}{\xm},
\eeq
with $\xm=1/\lambda$.

\vskip4pt\noindent\textbf{(b)}
Let us now consider a geometric distribution of points, which is the discrete version of the previous example.
Starting from some origin,
let us mark the integer points as renewal events with probability $p$.
Let $X$ be the number of unmarked points before the first marked point.
Its distribution is 
\be
f(k)=\prob(X=k)=p\,q^{k-1},
\ee
where $q=1-p$ and $k=1,2,\dots$.
Thus $\xm=1/p$.
Applying (\ref{eq:discretEquil}) leads to 
\beq\label{eq:geop0}
(p_0)_\eq(r)=q^r,
\eeq
which is simply the probability that $r$ consecutive points are unmarked.
Since $r$ is an integer, the analogue of (\ref{eq:micror}) can be obtained by taking the limit of (\ref{eq:geop0}) in the regime where $p=1/\xm$ is small.

\vskip4pt\noindent\textbf{(c)}
For a uniform distribution $\mathcal{U}(0,2d)$, such that $\xm=d$,
 (\ref{eq:equilib}) gives
\be
(p_{0})_{\eq}(r)=\bigac{1-\frac{r}{2d}}^2=1-\frac{r}{d}+\frac{r^2}{4d^2}.
\ee

\vskip4pt\noindent\textbf{(d)}
We now consider broad distributions of intervals (\ref{eq:powerlaw}) with tail index $\th>1$.
First, (\ref{eq:fEeq}) implies
\beq\label{fBEeq}
(f_{E})_\eq(\fo)
\stackunder{\approx}{\fo\rightarrow \infty}\frac{c}{\th\xm}\frac{1}{\fo^\th}.
\eeq
Then, in the stationary regime $1\ll r\ll j$, 
(\ref{eq:equilib}) yields
\beq\label{p0tt_eq}
(p_0)_\eq(r)\approx \frac{c}{\th(\theta-1)\xm}\,\frac{1}{r^{\th-1}}.
\eeq
In the scaling regime where $j$ and $r$ are large and
comparable, using (\ref{pntu}) and (\ref{fE}), we obtain
\be
\hat{p}_{0}(s,u)\approx \frac{a}{\xm}
\frac{s^{\theta -1}-u^{\theta -1}}{s(s-u)}, 
\ee
where $a$ is defined in (\ref{eq:adef}), which by inversion yields \cite{glrenew},
\beq\label{p0jr_l2}
p_{0}(j,r)\approx\frac{c}{\th(\theta -1)\xm}\left( \frac{1}{r^{\th-1}}-\frac{1}{(j+r)^{\th-1}}
\right) . 
\eeq
For $1\ll r\ll j$ we recover the equilibrium expression (\ref{p0tt_eq}), while for $1\ll j\ll r$
we obtain
\be
p_{0}(j,r)\approx
\frac{c}{\th\xm}
\frac{j}{r^{\theta}}.
\ee

\vskip4pt\noindent\textbf{(e)}
As can be seen on the three examples (a), (b), (c), whenever $r/\xm\ll1$, the hole probability $(p_0)_\eq(r)\approx 1-r/\xm$, which is the same behaviour as for equally spaced points (with $d=\xm$).
Example (d) does not allow to decide on the matter since only the tail of $f(k)$ is given.
Let us therefore consider an explicit example where the distribution $f(k)$ of the random spacing $X$ is known for all values of its argument.
We take $X=U^{-1/\th}$
where $U$ is a uniform random variable $\mathcal{U}(0,1)$, hence $X\ge1$.
So, for $k<1$, $f(k)=0$ and for $k\ge1$, ($k\in \mathbb{R}$),
\be
f(k)=\frac{\th}{k^{1+\th}},
\ee
with mean $\xm=\th/(\th-1)$.
Using (\ref{eq:equilib}), we get
\be
(p_{0})_{\eq}(r)=\cases{1-\frac{r}{\xm}&$0\le r\le 1$\\\frac{1}{\th\, r^{\th-1}}&$r\ge1$.}
\ee
This result confirms the general behaviour of $(p_{0})_{\eq}(r)$ for $r/\xm\ll1$.
The last line reproduces the right side of (\ref{p0tt_eq}).

\subsection{Broad distributions of intervals with index $\theta <1$}

For such distributions the system is never at equilibrium.
This is manifested by the fact that at long distances the system becomes self-similar.

\vskip8pt\noindent
\textit{Hole probability in the scaling regime}
\vskip4pt\noindent
We start by a reminder (see \cite{glrenew}).
In the scaling regime where $j$, $r$ and $\fo$ are large and comparable, we have, from (\ref{fE}) (see \ref{app:word} for notations),
\beq\label{fEscal}
\lim_{j\rightarrow \infty}f_{j^{-1}E}(x)=\frac{\sin \pi \theta}{\pi}
\frac{1}{x^{\theta}(1+x)}\qquad (0<x<\infty),
\eeq
thus, according to (\ref{p0jr}),
\begin{eqnarray}
p_{0}(j,r) &\approx&
\frac{\sin \pi \theta}{\pi}\int_{r/j}^{\infty}\dd x\,\frac{1}{x^{\theta}(1+x)}
=\int_{0}^{j/(j+r)}\dd x\,\beta _{\theta ,1-\theta}(x)
\nonumber \\
&=&\frac{\sin \pi \theta}{\pi}\mathrm{B}\left( \frac{j}{j+r};\th,1-\th\right)
\nonumber \\
&=&
1-\frac{\sin \pi \theta}{\pi}\textrm{B}\left( \frac{r}{j+r};1-\th,\th\right) ,
\label{eq:p0jrlt1}
\end{eqnarray}
where the beta distribution is defined as
\be
\beta _{a,b}(x)=\frac{\Gamma (a+b)}{\Gamma (a)\Gamma (b)}x^{a-1}(1-x)^{b-1} ,
\ee
and $\mathrm{B}(\cdot)$ is the incomplete beta function
\beq\label{eq:incomp}
\textrm{B}(z;a,b)=\int_0^z\dd x\,x^{a-1}(1-x)^{b-1}.
\eeq
For example, for $\th =\frac{1}{2}$, 
\beq\label{p0jrscal}
p_{0}(j,r)\approx1-\frac{2}{\pi}\arcsin \sqrt{\frac{r}{j+r}}.
\eeq
Moreover, in the same scaling regime $1\ll j\sim r$, the mean number of events occurring between $j$ and $j+r$ reads
\cite{glrenew},
\beq \label{Njrl1}
\mean{N(j,r)} \approx \frac{\sin \pi \th}{\pi c}\,((j+r)^{\th}-j^{\th}).
\eeq

A consequence of (\ref{eq:p0jrlt1}) and (\ref{Njrl1}) is that, in the regime of short separations ($1\ll r\ll j$), the probability of finding an event between $j$ and $j+r$ goes to zero, i.e., $p_0(j,r)\to1$.
In other words, in order to have a chance to
observe a renewal, one has to wait a duration $r$ of order $j$.
The intuitive explanation is that, as $j$ is growing, larger and larger
intervals of time $ k $ may appear. 
The density of events at large times is therefore decreasing
(see \cite[vol 1 p.~322]{feller}).
These observations can be made more precise by noting, firstly, that
in the regime of short separations ($1\ll r\ll j$) (\ref{eq:p0jrlt1}) yields
\beq\label{eq:freeshort}
p_0(j,r)\approx 1-\frac{\sin\pi\th}{\pi(1-\th)}\bigac{\frac{r}{j}}^{1-\th},
\eeq
and, secondly, that the mean density of events between $j$ and $j+r$ has, in the same regime, the expression
\beq\label{meanNoverr}
\frac{\mean{N(j,r)}}{r}\approx \frac{\th\sin\pi\th}{\pi c}\frac{1}{j^{1-\th}}.
\eeq

On the other hand, in the opposite regime of large separations between $j$ and 
$j+r$ ($1\ll j\ll r$), (\ref{eq:p0jrlt1}) yields the aging form 
\beq\label{eq:freelarge}
p_{0}(j,r)\approx
\frac{\sin \pi \theta}{\pi \theta}
\left( \frac{j}{r}\right)^{\theta}. 
\eeq
To summarise, 
\beqa\label{eq:web}
\lim_{r\to\infty}p_0(j,r)=0\quad \forall j\ \mathrm{finite},
\nonumber\\
\lim_{j\to\infty}p_0(j,r)=1\quad \forall r\ \mathrm{finite}.
\eeqa
This behaviour was dubbed \textit{weak ergodicity breaking} in \cite{bouchaud,dean}, in the sense that `true ergodicity breaking only sets in after infinite waiting time', i.e., $j\to\infty$ in the present context.

This ends the survey on the hole probability for free renewal processes.
We are now in position to derive the new result given in (\ref{eq:buffonfree}).

\vskip8pt\noindent
\textit{The Buffon probability}
\vskip4pt\noindent
Finally the answer to the Buffon problem consists in estimating the integral (\ref{eq:P0rLdef}) in the regime 
$1\ll r\sim L$.
Introducing the scaling variables 
\be
x=\frac{j}{L},\quad y=\frac{r}{L},
\ee
and using (\ref{eq:p0jrlt1}),
we obtain the scaling form 
\beq\label{eq:buffonfree0}
P_0(r,L)\approx C(y)=\frac{\sin \pi \theta}{\pi}
\frac{1}{1-y}\int_0^{1-y}\dd x\,
\mathrm{B}\left( \frac{x}{x+y};\th,1-\th\right),
\eeq
which, after some algebra, yields the universal result
\beq\label{eq:buffonfree}
C(y)=\frac{\sin \pi \theta}{\pi}\frac{1}{1-y}
\bigac{\mathrm{B}(1-y;\th,1-\th)-\frac{y^{1-\th}(1-y)^{\th}}{\th}}.
\eeq
For instance, for $\th=1/2$, we have
\be
C(y)=\frac{2}{\pi}\bigac{\frac{\arccos\sqrt{y}}{1-y}-\sqrt{\frac{y}{1-y}} }.
\ee
For $y\to0$, (\ref{eq:buffonfree}) yields
\be
C(y)\approx 1-\frac{\sin\pi\th}{\pi \th(1-\th)}y^{1-\th},
\ee
while for $y\to1$, it yields
\be
C(y)\approx \frac{\sin\pi\th}{\pi}\frac{(1-y)^{\th}}{\th(1+\th)}.
\ee
Both limit expressions could have been obtained by integrations of the limit expressions of $p_0(j,r)$ in the same regime, i.e., respectively (\ref{eq:freeshort}) and (\ref{eq:freelarge}).

\section{Probability of an empty interval for a tied-down renewal process}
\label{sec:p0TD}

We now consider the tied-down renewal process, with condition (\ref{eq:condition}) at $L$.
Consider the interval of length $r$, with left endpoint located on site $j$ (see figure \ref{fig:definition-TD}).
The aim of this section is to give the general expression of the probability $p_0(j,r,L)$ for the interval $(j,j+r)$ to be empty, which is the quantity of central interest.
Here we shall describe the process using a discrete formalism (see also \cite{jsp}).
This framework complements and enriches the viewpoint based on the continuum formalism used previously in \cite{wendel2,wendel1}. It is also appropriate for the description of the tied-down random walk.

\subsection{Weight of configurations}
\label{sec:weight}
A configuration of the process, $\cc=\{X_1=k_1,X_2=k_2,\dots,X_{N_L}=k_n,N_L=n\}$, has weight \cite{wendel2,wendel1,jsp}
\beq\label{eq:cc}
\prob(\cc)=\frac{1}{Z(L)}f(k_1)\dots f(k_n)\,\delta\Big(\sum_{i=1}^n k_i,L\Big),
\eeq
where the partition function,
\beqa\label{eq:ZwL}
Z(L)&=&\sum_{\cc}\prob(\cc)
=\sum_{n\ge0}\sum_{\{k_i\}} f(k_1)\dots f(k_n)\delta\Big(\sum_{i=1}^n k_i,L\Big)
\nonumber\\
&=&\sum_{n\ge0}\prob(S_{n}=L)=\prob(S_{N_L}=L)=\mean{\delta(S_{N_L},L)},
\eeqa
is the probability that a renewal occurs at $L$ (with the notation $S_n=X_1+\cdots+X_n$, see (\ref{eq:Sn})).
Its generating function ensues from (\ref{eq:ZwL}),
\beqa\label{eq:Zz}
\tilde Z(z)&=&\sum_{L\ge0}z^L Z(L)=\sum_{n\ge0}\sum_{\{k_i\}} z^{k_1}f(k_1)\dots z^{k_n}f(k_n)
\nonumber\\
&=&\sum_{n\ge0}\tilde f(z)^n=\frac{1}{1-\tilde f(z)},
\eeqa
with
\be
\tilde f(z)=\sum_{k\ge1}z^kf(k).
\ee
Consider for instance the case where
\beq
\label{app:ftilde}
\tilde f(z)=1-\sqrt{1-z},
\eeq
corresponding to
\beq\label{eq:first}
f(k)=\frac{1}{2^{2 k-1}}\frac{(2 k-2)!}{( k-1)! k!}\stackunder{\approx}{k\to\infty} \frac{1}{2\sqrt{\pi}k^{3/2}}.
\eeq
Thus
\beq\label{app:utilde}
\tilde Z(z)=\frac{1}{\sqrt{1-z}},
\eeq
and
\beq\label{eq:return}
Z(L)=\frac{1}{2^{2L}}\binom{2L}{L}\stackunder{\approx}{L\to\infty}\frac{1}{\sqrt{\pi L}}.
\eeq
The first values of these quantities read $Z(0)=1,Z(1)=1/2,\dots$, $f(1)=1/2, f(2)=1/8,\dots$.
These quantities have a simple interpretation for the tied-down random walk \cite{feller,wendel2,wendel,wendel1,jsp}.
The partition function $Z(L)$ corresponds to
the probability of return at the origin of the walk at time $2L$,
while $f(k)$ is the probability of first return at the origin of the walk at time $2k$.

\subsection{Probability for the interval $(j,j+r)$ to be empty}
The general expression of the probability $p_0(j,r,L)$ for the interval $(j,j+r)$ to be empty reads
\beq\label{eq:GB}
p_0(j,r,L)=\frac{1}{Z(L)}\sum_{k_1=0}^{j}\,\sum_{k_2=j+r+1}^L
Z(k_1)f(k_2-k_1)Z(L-k_2).
\eeq
This expression can easily be understood.
The summand $Z(k_1)f(k_2-k_1)Z(L-k_2)$ is the probability to find a renewal on the left of the needle ($k_1\le j$), followed by an empty interval $k_2-k_1$, where $k_2$ is the first renewal encountered on the right of the needle ($k_2>j+r$)%
\footnote{We count a renewal located at $j$, i.e., on the left end of the needle, as being outside the needle, while a renewal located at $j+r$, the right end, is counted as being inside.
This convention has no consequence for the sequel.}.
This expression is left-right symmetric, as can be seen by changing $k$ to $L-k$.

In order to answer the Buffon problem, we have to compute the probability $P_0(r,L)$ for an interval of length $r$ placed uniformly at random on $(0,L)$ to be empty.
This Buffon probability is obtained by summation of (\ref{eq:GB}) on all the possible positions of the left end of the needle, $0\le j\le L-r$,
so
\beq\label{eq:GB2}
P_0(r,L)=\frac{1}{L-r+1}\sum_{j=0}^{L-r-1}p_0(j,r,L).
\eeq
The upper bound $L-r-1$ takes into account the fact that $p_0(j,r,L)$ vanishes for $j\ge L-r$ since there is a renewal at $j=L$.

\subsection{Probability of an empty interval at equilibrium}
\label{sec:equilibriumTD}
We first illustrate (\ref{eq:GB}) and (\ref{eq:GB2}) on two examples of processes converging to equilibrium.

\vskip4pt\noindent
\noindent\textbf{(a)}
Consider first
the simple example of the geometric distribution of points defined in \S \ref{sec:equilibrium}.
Simple calculations lead to
\be
\prob(S_n=L)=\binom{L-1}{n-1}p^nq^{L-n},
\ee
so
\be
Z(L)=\sum_{n\ge0}\prob(S_n=L)=p,
\ee
as it should, since, as said above, $Z(L)$ is the probability of finding a renewal (i.e., a marked point) at $L$.
As a result of (\ref{eq:GB}), one gets
\be
p_0(j,r,L)=(p_0)_\eq(r)=q^r
\qquad (0\le j\le L-r-1),
\ee
in accord with (\ref{eq:geop0}).
Hence
\beq\label{eq:geoP0}
P_0(r,L)=\frac{L-r}{L-r+1}q^r\stackunder{\longrightarrow}{L\to\infty}P_0(r)=q^r.
\eeq
The constant value of $Z(L)=1/\xm=p$ is in line with the fact that the process is at equilibrium.

\vskip4pt\noindent\textbf{(b)}
Consider now a distribution of the intervals $X_1,X_2,\dots$ with power-law tail (\ref{eq:powerlaw}) and index $\th>1$, the first moment $\xm$ of the distribution $f(k)$ is finite, and (see, e.g., \cite{wendel1,jsp}),
\beq\label{eq:ZLgt1}
Z(L)\stackunder{\approx}{L\to\infty} \frac{1}{\xm}+\frac{c}{\th(\th-1)\xm^2}L^{1-\th},
\eeq
which shows that asymptotically the process becomes stationary, with a density of renewals equal to $1/\xm$ at large $L$.
Keeping only the dominant term in (\ref{eq:ZLgt1}) yields
(\ref{p0jr_l2}) once again, then by summation on $j$ according to (\ref{eq:GB2}) we obtain
\be
P_0(r,L) \stackunder{\approx}{L\to\infty}P_0(r)= (p_0)_\eq(r),
\ee
a result in accord with (\ref{p0tt_eq}).

\subsection{Asymptotic analysis of the hole probability when $\th<1$ }
\label{sec:scaling}

This section contains the main results of the present work.
As for free renewal processes, when the distribution (\ref{eq:powerlaw}) has tail index $\th<1$, the system is never at equilibrium
and becomes self-similar at long distances.
The scaling analysis of (\ref{eq:GB}) proceeds as follows.

If $\th<1$,
it is known, and easy to infer from (\ref{eq:Zz}), that (see, e.g., \cite{wendel1,jsp}),
\beq\label{eq:ZLlt1}
Z(L)\stackunder{\approx}{L\to\infty} \frac{\th\sin\pi\th}{\pi c}\frac{1}{L^{1-\th}},
\eeq
which means that renewals become rarer and rarer as $L$ increases (their density decays as $1/L^{1-\th}$)---a manifestation of the non-stationarity of the process.
Note that (\ref{eq:ZLlt1}) is the same as (\ref{meanNoverr}) on replacing $j$ by $L$.
Equations (\ref{eq:first}) and (\ref{eq:return}) are particular cases of (\ref{eq:powerlaw}) and (\ref{eq:ZLlt1}).

In the continuum limit, with $1\ll j\sim r\sim L$, (\ref{eq:GB}) can be recast as 
\bea
p_0(j,r,L)\approx\frac{1}{Z(L)}\int_{0}^{j}\dd k_1\,Z(k_1)\int_{j+r}^L\dd k_2\,f(k_2-k_1)Z(L-k_2).
\eea
Introducing the scaling variables 
\be
x=\frac{j}{L},\quad y=\frac{r}{L},\quad a=\frac{k_1}{L},\quad b=\frac{k_2}{L},
\ee
and using (\ref{eq:ZLlt1}), the probability of interest takes the scaling form
\be
p_0(j,r,L)\approx c(x,y),
\ee
where
\beq
 c(x,y)= \frac{\th\sin\pi\th}{\pi}
\int_0^x\dd a
\int_{x+y}^1
\dd b\,\frac{1}{a^{1-\th}(b-a)^{1+\th}(1-b)^{1-\th}}.
\eeq
The second integral can be performed explicitly, 
\beq
\int_{z}^1
\dd b\,\frac{1}{(b-a)^{1+\th}(1-b)^{1-\th}}=\frac{(1-z)^\th}{\th(1-a)(z-a)^\th}.
\eeq
We are thus left with
\beq\label{eq:fctnScal}
 c(x,y)=\frac{\sin\pi\th}{\pi}(1-x-y)^\th\int_0^x\dd a\,\frac{1}{(1-a)a^{1-\th}(x+y-a)^\th}.
\eeq
For $y=0$ this expression gives $c(x,0)=1$, as it should.
Making the following change of variable in (\ref{eq:fctnScal}),
\be
t=\frac{x+y-a}{(1-a)(x+y)},
\ee
yields the more explicit expression
\beq\label{eq:scalingfctn}
c(x,y)=1-\frac{\sin\pi\theta}{\pi}\mathrm{B}\left(\frac{y}{(1-x)(x+y)};1-\th,\th\right),
\eeq
where $\mathrm{B}(\cdot)$ is the incomplete beta function (see (\ref{eq:incomp})),
\be
\textrm{B}(z;1-\th,\th)=\int_0^z\dd x\,x^{-\th}(1-x)^{\th-1}.
\ee
This expression---which is the main result of this section---is universal since it only depends on the tail index $\th$ and not on the details of the distribution $f(k)$.
Remarkably, it only depends on a single variable, namely the cross-ratio of the four points $x_1=0$, $x_2=x$, $x_3=x+y$ and $x_4=1$,
\be
\zeta=\frac{(x_1-x_4)(x_2-x_3)}{(x_1-x_3)(x_2-x_4)}= \frac{y}{(1-x)(x+y)}.
\ee
For $\th=1/2$, (\ref{eq:scalingfctn}) reduces to the known result for the Brownian bridge (see (\ref{eq:pont})),
\beq\label{eq:CBb}
 c^{\bb}(x,y)=1-\frac{2}{\pi}\arccos\sqrt{\frac{x(1-x-y)}{(1-x)(x+y)}},
\eeq
which satisfies $\cos(\pi c^{\bb}/2)^2=\zeta$.

In the regime of short separations, such that $1\ll r\ll j\sim L$, hence $y\to0$, the expression (\ref{eq:scalingfctn}) simplifies to
\beq\label{eq:Cxyshort}
c(x,y)\approx 1-\frac{\sin\pi\theta}{\pi(1-\theta)}\left(\frac{y}{x(1-x)}\right)^{1-\th}.
\eeq
In the regime of large separations, such that $1\ll j\ll r\sim L$, hence $x\to0$, the expression (\ref{eq:scalingfctn}) gives
\beq\label{eq:Cxylarge}
c(x,y)\approx \frac{\sin\pi\th}{\pi\th}\left(\frac{x(1-y)}{y}\right)^\th.
\eeq
These two last results were found by another method in \cite{wendel2} (see the remark in section \ref{sec:variant}). 

\textit{Remark.}\;
The counterpart of (\ref{eq:scalingfctn}) for the free renewal process
is given by (\ref{eq:p0jrlt1}).
The latter is recovered by taking (\ref{eq:scalingfctn}) in the regime $1\ll j\sim r\ll L$.
Accordingly, (\ref{eq:Cxyshort}) and (\ref{eq:Cxylarge}) generalise (\ref{eq:freeshort}) and (\ref{eq:freelarge}).
The property of weak ergodicity breaking (\ref{eq:web}) still holds in the present situation.

The last step consists in deriving the integrated scaling function.
By integration of (\ref{eq:scalingfctn}) on $x$, 
the probability (\ref{eq:GB2}) takes the scaling form
\be
P_0(r,L)\approx C\bigac{y=\frac{r}{L}}=\frac{1}{1-y}\int_0^{1-y}\dd x \,c(x,y),
\ee
yielding, after some algebra,
\beq\label{eq:scaling1}
C(y)=\frac{\sin \pi\th\,\Gamma(\th)}{2\sqrt{\pi}\,\Gamma(1/2+\th)}\frac{1+y}{1-y}\,
\mathrm{B}\!\left(\frac{(1-y)^2}{(1+y)^2};\frac{1}{2}+\th,1-\th\right).
\eeq
This expression gives the answer to the Buffon problem for distributions with power-law tails of index $\th<1$.
It is universal, as was (\ref{eq:scalingfctn}).

In the regime of short separations, $y\to0$, (\ref{eq:scaling1}) yields
\beq\label{eq:Cbary}
 C(y)\approx
1-\frac{\Gamma(\th)}{\Gamma(2\th)\Gamma(2-\th)}
y^{1-\theta},
\eeq
which can also be obtained by integration of (\ref{eq:Cxyshort}) on $x$ \cite{arxiv,comment}.
In the other regime of interest, $y\to1$ (large separations), we have
\beq
 C(y)\approx \frac{(1-\th)\Gamma(\th)}{2(1+\th)\Gamma(2\th)\Gamma(2-\th)}(1-y)^{2\th}.
\eeq 
Finally, for $\th=1/2$, the expression (\ref{eq:scaling1}) simplifies to
\beq\label{eq:BBCy}
 C(y)=\frac{1-\sqrt{y}}{1+\sqrt{y}},
\eeq
which coincides with the result given in (\ref{eq:p0rL}) for $C^{\bb}(y)$, as it should.

\subsection{A variant}
\label{sec:variant}

The probability of an empty interval $p_0(j,r,L)$ can also be expressed in terms of the distribution 
$p_E(j,\fo,L)=\prob(E_j=\fo)$ of the forward recurrence-time (or excess time) $E_j$, starting at $j$,
(see figure \ref{fig:figEAS}).
Indeed,
\beq\label{eq:p0pE}
p_0(j,r,L)=\prob(E_j>r)
=\sum_{\fo=r+1}^{L-j}p_E(j,\fo,L).
\eeq
Comparing to (\ref{eq:GB}), we have, setting $k_2-j=\fo$,
\beq\label{eq:pEjlL}
p_E(j,\fo,L)=\frac{Z(L-j-\fo)}{Z(L)}\sum_{k_1\le j}Z(k_1)f(j+\fo-k_1).
\eeq
The scaling analysis of this quantity---still focusing on the case where $\th<1$---leads to 
\beq\label{eq:pEjeL}
p_E(j,\fo,L)\approx \frac{1}{L}g_E(x,\xi),
\eeq
with $x=j/L$ and $\xi=\fo/L$ and where
\beqa\label{eq:pEjeLas}
g_E(x,\xi)&=&\frac{\th\sin\pi\th}{\pi}
\int_0^x\dd a\,\frac{(1-x-\xi)^{\th-1}}{a^{1-\th}(x+\xi-a)^{1+\th}}
\nonumber\\
&=&\frac{\sin\pi\th}{\pi}
\frac{x^\th}{(x+\xi)\xi^\th(1-x-\xi)^{1-\th}}.
\eeqa
The scaling function $g_E(x,\xi)$ is a generalisation to the tied-down process of the corresponding scaling function for the free renewal process given in (6.6) of \cite{glrenew}.
The latter is recovered by letting $L\to\infty$ in (\ref{eq:pEjeLas}).
Summing (\ref{eq:pEjeLas}) upon $\xi$ in the interval $(y,1-x)$ gives back (\ref{eq:scalingfctn}).

\subsection*{Remark}
The numerator of the generating function of $p_E(j,\fo,L)$ with respect to all its arguments (with conjugate variables $u,v,z$) is given by
\beqa
\tilde p_E(u,v,z)\num&=&\sum_{L\ge0} z^L \sum_{j=0}^L u^j\sum_{\fo=1}^L v^\fo p_E(j,\fo,L)\num
\nonumber\\
&=&\tilde Z(z)\tilde Z(zu)\frac{v\tilde f(zv)-u\tilde f(zu)}{v-u}.
\eeqa
This expression is the discrete counterpart of the expression given in \cite{wendel2} for the density of the forward recurrence-time in Laplace space.
According to (\ref{eq:p0pE}) it is related to the generating function of $p_0(j,r,L)$ with respect to its arguments $j,r,L$ (with conjugate variables $u,v,z$) as
\beqa\label{eq:fctngenp0}
\tilde p_0(u,v,z)\num
&=&\sum_{L\ge0} z^L \sum_{j=0}^L u^j\sum_{r=0}^L v^r p_0(j,r,L)\num
\nonumber\\
&=&\frac{\tilde p_E(u,1,z)\num-\tilde p_E(u,v,z)\num}{1-v}.
\eeqa
The asymptotic analysis of this expression was made in \cite{wendel2} in the regimes of short and large separations,
yielding respectively (\ref{eq:Cxyshort}) and (\ref{eq:Cxylarge}).
The direct approach based on (\ref{eq:pEjlL}) is more convenient for the derivation of the complete expression of the scaling function (\ref{eq:scalingfctn}), valid in all regimes.

\section{Complementary remarks}
\label{sec:complement}

This section adds some complements to the body of the text.

\subsection{Straddling interval and inspection paradox}
\label{sec:stradl}

In Laplace space, for a free renewal process, the density of the interval straddling $j$ (\ref{eq:straddef}) reads
\beq\label{eq:stradLap}
\lap{j,k} f_{X_{N_j+1}}(j,k)=\hat f_{X_{N_j+1}}(s,u)=\frac{1}{1-\hat f(s)}\frac{\hat f(u)-\hat f(s+u)}{s}.
\eeq
This expression can easily be demonstrated by the methods of \cite{glrenew}.
We have
\be
f_{X_{N_j+1},N_j}(j,k,n)=\mean{\delta(k-S_{n+1}+S_n))I(S_n<j<S_{j+1})},
\ee
where $I(\cdot)$ is the indicator function of the event inside the parenthesis.
Laplace transforming with respect to $j$ and $k$ and summing upon $n$ leads to (\ref{eq:stradLap}).

Hence, at equilibrium, we have, using (\ref{eq:stradLap}),
\be
(\hat f_{X_{N_j+1}})_\eq(u)=\lim_{s\to0}s\hat f_{ X_{N_j+1}}(s,u)
=-\frac{1}{\xm}\frac{\dd\hat f(u)}{\dd u},
\ee
we thus recover (\ref{eq:straddle}),
\be
(f_{ X_{N_t+1}})_\eq(k)=\frac{kf(k)}{\xm}.
\ee
The inspection paradox, which states that $\mean{X_{N_j+1}}$ is larger than $\xm$, ensues immediately.
Indeed, if $f(k)$ is a narrow distribution or a broad distribution with tail index $\th>2$, one has, at equilibrium,
\beq\label{eq:tauN1}
\mean{X_{N_j+1}}_\eq=\frac{1}{\xm}\int_0^\infty\dd k\, k^2\,f( k)= \frac{\mean{X^2}}{\xm}>\xm.
\eeq
In contrast, if $1<\th<2$, this second moment diverges.
The inspection paradox is all the more true.
The average straddling interval diverges as \cite{glrenew}
\be
\mean{X_{N_j+1}}\stackunder{\approx}{j\to\infty} \frac{c}{(\th-1)(2-\th)\xm}j^{2-\th}.
\ee

\subsection{Behaviour of the hole probability $p_0$ when $r\to0$}
It was noted on examples in \S \ref{sec:equilibrium} that for microscopic values of $r$ the behaviour of the hole probability at equilibrium is given by $(p_0)_\eq(r)\approx 1-r/\xm$.
Let us prove this result in all generality and extend it to the case of broad laws with tail index $\th<1$.

When $r$ is microscopic the asymptotic behaviours of the free and tied-down processes are the same, so we can 
restrict the discussion to the case of the free process.
We use (\ref{pntu}) and (\ref{fE}).
If $r$ is microscopic, the conjugate Laplace variable $u$ is large, 
while for large $j$, the conjugate Laplace variable $s\to0$.
Therefore, for narrow distributions or broad laws with tail index $\th>1$,
we have
\be
\hat{f}_{E}(s,u) \approx \frac{1}{u}\frac{1}{s\xm},
\ee
which, by Laplace inversion, yields
\beq\label{petit-r}
(p_0)_\eq(r)\approx 1-\frac{r}{\xm},
\eeq
which confirms the generality of the results found in \S \ref{sec:equilibrium}.
On the other hand, for broad laws with tail index $\th<1$, the same reasoning yields
\be
\hat{f}_{E}(s,u) \approx \frac{1}{u}\frac{1}{as^\th},
\ee
thus, in the regime $1\sim r\ll j$,
\be
p_0(j,r)\approx 1-\frac{r}{\mean{X_j}},
\ee
where we introduced the notation $\mean{X_j}$ for the typical interval length 
\be
\mean{X_j}\approx\frac{\pi c}{\th\sin\pi\th}j^{1-\th}.
\ee
The beauty of this result is that one recognises the asymptotic expression of $1/Z(j)$ (see (\ref{eq:ZLlt1})).
Now, as seen in (\ref{eq:ZLgt1}), for distributions with a finite first moment,  the estimate $Z(j)\approx 1/\xm$ holds for $j$ large.
The present discussion extends this result to the case of distributions without finite first moment, i.e., asymptotically, $Z(j)$ has the interpretation of the inverse typical interval length $\mean{X_j}$.
This interpretation was actually already pointed out in the comment below (\ref{eq:ZLlt1}) in view of
(\ref{meanNoverr}).

Let us finally mention that (\ref{petit-r}) is more general and also holds if the intervals $X_1,X_2,\cdots$ are not independent (see \cite{mehta}).

\subsection{Buffon problem on a ring}
Instead of considering the probability of an empty interval on a segment $(0,L)$, this probability can also be considered on a ring of size $L$.
However this requires giving a precise definition to the process.
If one keeps the definition of the configurations of the process given in \S \ref{sec:weight}, with weights (\ref{eq:cc}), meaning that the two endpoints 0 and $L$ are identified into a single point,
the only change to be made in the definition of the Buffon probability given by (\ref{eq:GB2}) is the interval of integration.
The latter is changed into
\beq\label{eq:ring}
P_0(r,L)=\frac{1}{L}\sum_{j=0}^{L-r-1}p_0(j,r,L),
\eeq
or into
\beq\label{eq:ring+}
P_0(r,L)=\frac{1}{L}\int_{0}^{L-r}\dd j\,p_0(j,r,L),
\eeq
in the continuum formalism.

\subsection{Correlation function}
We defined the probability of having $n$ renewals in the interval $(j,j+r)$ as, (see (\ref{eq:pnjr})),
\be
p_n(j,r,L)=\prob(N(j,r)=n),
\ee
where $N(j,r)$ is the number of renewals in $(j,j+r)$.
Knowing this probability allows the determination of the correlation function
\beq\label{eq:correl}
\langle (-1)^{N(j,r)}\rangle
=\sum_{n\ge0}(-1)^n
p_n(j,r,L).
\eeq
This correlation is a natural quantity to consider if the intervals represent $\pm1$ spin domains.
The study of this spin-spin correlation function was performed in \cite{glrenew} for the case of a semi-infinite one-dimensional system, that is, for a free renewal process as in figure \ref{fig:definition-libre}.
Reference \cite{wendel2} gave an extension of this former study to the case of a system of fixed length $L$, conditioned by (\ref{eq:condition}), that is, for a tied-down renewal process as in figure \ref{fig:definition-TD}.

It turns out that, for both free and tied-down renewal processes, when the distribution $f(k)$ is of the form (\ref{eq:powerlaw}), with $0<\th<2$, this correlation function is asymptotically dominated by its first term, namely $p_0(j,r)$ or $p_0(j,r,L)$ respectively,
and therefore the knowledge of this hole probability suffices to determine the correlation (\ref{eq:correl}) \cite{glrenew, wendel2}.

\section{Discussion}
\label{sec:discussion}

In this paper the focus was on the probability of an empty interval---or hole probability---both for free and tied-down renewal processes.
In the first case the hole probability $p_0(j,r)$ is a function of 
$j$, the position of the left end of the needle, and $r$, the length of the needle, while in the second case $p_0(j,r,L)$ also depends on the length $L$ of the system.

The present work is a completion of \cite{glrenew} and \cite{wendel2}.
The main novelty with respect to \cite{glrenew} 
is the determination of the Buffon probability  (see (\ref{eq:buffonfree0}) and (\ref{eq:buffonfree}))
for a broad distribution of intervals $f(k)$ with a power-law tail (\ref{eq:powerlaw}) of index $\th<1$
in the scaling regime $1\ll r\sim L$.
As for \cite{wendel2}, the main novelties are the complete expression (\ref{eq:scalingfctn}) of the universal scaling function $c(x,y)$, as well as its integrated expression, the Buffon probability (\ref{eq:scaling1}), in the regime where all the arguments of 
$p_0(j,r,L)\approx c(x=j/L,y=r/L)$ are large and comparable, again for broad laws with tail index $\th<1$.
The results given in \cite{wendel2} were limited to the expressions of $c(x,y)$ in the two regimes of short and large separations, i.e., respectively (\ref{eq:Cxyshort}) and (\ref{eq:Cxylarge}) and to the expression (\ref{eq:CBb}) of $c(x,y)$ for the case of the Brownian bridge.
For a large system, the expression
(\ref{eq:scalingfctn}) still keeps a dependence in the ratio $x=j/L$, which demonstrates the nonstationarity of the process.
In contrast, if $\th>1$, the process becomes stationary, when $j\sim L\to\infty$, i.e., no longer depends on $j$.

As mentioned in the Introduction, this kind of systems, made of a random number of iid intervals conditioned by the value of their sum, are naturally encountered in statistical physics,
e.g., the tied-down random walk \cite{wendel,wendel1}, the model introduced in \cite{burda2,bar2}, or more generally the \textit{linear models} defined by Fisher \cite{fisher}, such as the Poland-Scherraga model \cite{poland,poland2}, interface models or models of wetting \cite{gia1,gia2}.

The critical spin-spin correlation function of the model considered in \cite{bar2}, 
where the iid intervals $X_1,X_2,\dots$ represent positive or negative spin domains partitioning the segment $(0,L)$, 
with a power-law distribution (\ref{eq:powerlaw}) and tail index $\th<1$, 
identifies to (\ref{eq:correl}) \cite{comment}.
Its expression in the scaling regime is therefore given by (\ref{eq:scalingfctn}), or by (\ref{eq:scaling1}) after summation on $x$.
The computation of this correlation function, restricted to the regime of short separations,
was revisited in \cite{bmm,bmm1}, using heuristic methods which boil down to considering the system at equilibrium.
It was however pointed out in \cite{arxiv,comment} that applying
the equilibrium formalism of \S \ref{sec:equilibrium} or \S \ref{sec:equilibriumTD} to a situation where the tail exponent $\th<1$ is incorrect,
irrespective of whether the geometry of the system is a line segment \cite{bmm1} or a ring \cite{bmm}.%
\footnote{Taking periodic boundary conditions in the context of a spin model with Boltzmann weights given by (\ref{eq:cc}), as done in \cite{bmm}, brings the additional difficulty of having
to select configurations such that the parity of the number of domains is compatible with periodicity.
This issue is not addressed in \cite{bmm}.}

\ack It is a pleasure to thank G Giacomin and J M Luck for enlightening discussions and P Calka for useful correspondence.

\appendix

\section{A word on notations}
\label{app:word}

\noindent \textit{Asymptotic equivalence}

The symbol $\approx$ stands for asymptotic equivalence;
the symbol $\sim$ is weaker and means `of the order of'.

\medskip
\noindent \textit{Probability densities, Laplace transforms, limiting distributions}

The probability density function of the continuous random variable $X$ is denoted by $f_X(x)$, with
\be
f_X(x)=\frac{\dd}{\dd x}\prob(X<x).
\ee
In the case of the forward recurrence-time $E_j$,
the density
\be
f_{E}(j,\fo)=\frac{\dd }{\dd \fo}\prob(E_j<\fo)
\ee
has a dependence in the (time or space) coordinate $j$. 
Its Laplace transform with respect to $\fo$ is
\be
\hat{f}_{E}(j,u)=\lap{\fo}f_{E}(j,\fo)
=\left\langle {\e}^{-uE_j}\right\rangle =\int_{0}^{\infty}\dd \fo\,{\e}^{-u\fo}\,f_{E}(j,\fo), 
\ee
and its double Laplace transform with respect to $j$ and $\fo$
is denoted by 
\be
\fl
\hat{f}_{E}(s,u)=\lap{j,\fo}f_{E}(j,\fo)=\lap{j}\left\langle {\e}^{-uE_j}\right\rangle
=\int_{0}^{\infty}\dd j\,{\e}^{-sj}\int_{0}^{\infty}\dd \fo\,{\e}^{-u\fo}\,f_{E}(j,\fo).
\ee
When $\th<1$, $E_j$ scales 
asymptotically as $j$.
As $j\to\infty$ the rescaled variable $E_j/j$
 converges to a limit, denoted by 
\be
X=\lim_{j\to\infty}j^{-1}E_j,
\ee
with limiting density $f_X(x)$, (see (\ref{fEscal})).

\section{Correlation function for the Brownian bridge}
\begin{figure}[!ht]
\begin{center}
\includegraphics[angle=0,width=.9\linewidth]{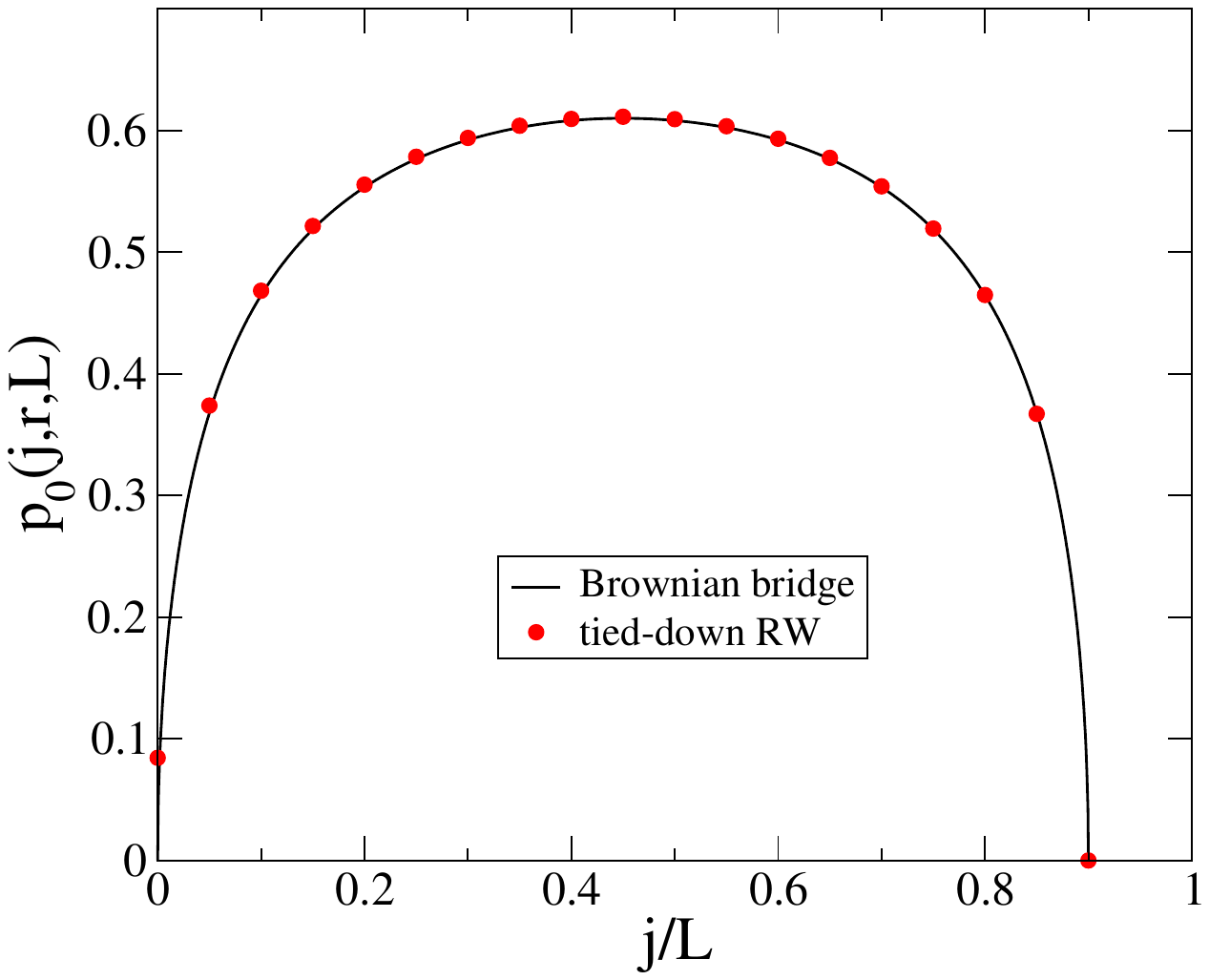}
\caption
{Exact probability $p_0(j,r,L)$ given by (\ref{eq:GB}) as a function of $j/L$ for the tied-down random walk (red dots). 
This probability is compared to the analytical expression (\ref{eq:pont}) of the correlation function of the sign of the position for the Brownian bridge (continuous curve).
Here $r=40$, $L=400$.
}
\label{fig:p0}
\end{center}
\end{figure}

This appendix is a reminder of results given in \cite{wendel2,arxiv}.
The continuum limit of the tied-down random walk defined in \S \ref{sec:definition} (see (\ref{eq:return}), (\ref{eq:first})) is the Brownian bridge.
A direct computation, based on the fact that the Brownian bridge is a Gaussian process, gives the correlation function defined in (\ref{eq:correl}) \cite{wendel2},
\beq\label{eq:pont}
\langle (-1)^{N(j,r)}\rangle=1-\frac{2}{\pi}\arccos\sqrt{\frac{j(L-j-r)}{(j+r)(L-j)}},
\eeq
yielding (\ref{eq:CBb}) for $c^{\bb}(x,y)$.
A comparison between (\ref{eq:GB}) for the tied-down random walk 
and (\ref{eq:pont}) for $L=400,r=40$ is given in figure \ref{fig:p0}, with excellent agreement.
This figure gives an illustration of the asymptotic dominance of $p_0(j,r,L)$ mentioned in \S \ref{sec:discussion}.

Integrating 
(\ref{eq:pont}) on $j$ gives
\beq\label{eq:p0rL}
C^{\bb}(y)=\frac{1-\sqrt{y}}{1+\sqrt{y}},
\eeq
in agreement with (\ref{eq:BBCy}) which was obtained from (\ref{eq:scaling1}) for $\th=1/2$.

\end{document}